\documentclass[prd,nofootinbib,english]{revtex4}

\usepackage{graphicx,float}
\usepackage{amsmath,amssymb,amsfonts}
\usepackage{epsfig,color}
\usepackage[thinlines]{easytable}
\usepackage{pdfpages}
\usepackage{array}
\usepackage{cancel}
\usepackage{mathtools}
\usepackage{accents}
\usepackage{subfigure}
\usepackage{enumitem}
\usepackage[dvipsnames]{xcolor}
 \usepackage{hyperref}
\hypersetup{
    colorlinks=true,
    linkcolor=blue,
    filecolor=magenta,      
   citecolor=blue
}

\begin{document}

\title{Conformal transformation of f(Q) gravity and its cosmological perturbations}

\author{Dehao Zhao}
\email{dhzhao@ucas.ac.cn}
\affiliation{School of Physical Sciences, University of Chinese	Academy of Sciences, Beijing 100049, China}

\begin{abstract}
Symmetric teleparallel gravity (STG) is a gravity theory which takes non-metricity tensor to describe gravity effects. In the STG framework, we study the conformal equivalent scalar-tensor theory of $f(Q)$ model and calculate the cosmological linear perturbations of the conformal transformed action. We confirm the result already present in references that $f(Q)$ gravity shows different degrees of freedom on different backgrounds at linear perturbation level. We also explain that this situation often means the linear perturbation theory breaks down and the model may suffer from strong coupling problem.
\end{abstract}

\maketitle

\section{Introduction}

Although General Relativity (GR) has achieved great success, it faces the difficulties in both theoretically and observationally, such as singularity of the universe, dark matter and dark energy. Therefore, GR is now generally considered as an effective theory and thought to be modified at high energy and/or large length scales. One of the simplest modifications to GR is the $f(R)$ gravity in which the Lagrangian density ($R$) of GR is replaced by its arbitrary function $f(R)$ \cite{Sotiriou:2008rp,DeFelice:2010aj}. There are two major reasons to explain why people study $f(R)$ gravity. First of all, $f(R)$ gravity is sufficient enough to capture some of the basic characteristics of
higher-order gravity, such as considering $f(R)$ as a series expansion
\begin{eqnarray}
	f(R)=\ldots+\frac{\alpha_2}{R^2}+\frac{\alpha_1}{R}-2 \Lambda+R+\frac{R^2}{\beta_2}+\frac{R^3}{\beta_3}+\ldots,
\end{eqnarray}
where $\alpha$, $\beta$ are coefficients and have suitable dimensions. Second, it is well-known that modified gravity theories in Riemann geometry in which gravity effects are described by the Riemann curvature, generally suffer from the so-called Ostrogradski instability \cite{Woodard:2006nt,Woodard:2015zca} or higher-order derivative instability, because the Riemann curvature hides second time derivative of metric. But $f(R)$ gravity can avoid this type of instability. This result can be easily understand by the conformal transformed action of $f(R)$ gravity which is just a Einstein-Hilbert action plus a minimally coupled scalar field, and since conformal transformation does not change the ghost behavior of theories, the $f(R)$ gravity is ghost-free. 

Since in the framework of Riemann geometry, it is difficult to modify gravity to avoid higher-order derivative instability, then we can do it in non-Riemann geometry. In this paper, we will consider the so-called symmetric teleparallel gravity (STG) \cite{Nester:1998mp,BeltranJimenez:2017tkd} framework. STG theory can be considered as a constrained metric-affine theory, and the quantity which describes gravity effects is non-metricity tensor, $Q_{\alpha\mu\nu}=\nabla_{\alpha}g_{\mu\nu}$, where $\nabla$ is covariant derivative which satisfies curvature-free and torsion-free conditions. Non-metricity tensor measures the failure of the affine connection to be metric-compatible. The STG model we are most concerned about is the so-called symmetric teleparallel equivalent of general relativity (STEGR), which is equivalent to GR and its action is equal to the Einstein-Hilbert action up to a surface term. The Lagrangian density of STEGR can be written as scalar $Q$, then a straightforward modification of STEGR is just to replace the scalar $Q$ to its arbitrary function $f(Q)$ \cite{Lu:2019hra,BeltranJimenez:2019tme}. Meanwhile, because non-metricity tensor only contains first-derivative of metric, the field equations will always be second order equations, remarkably simpler than $f(R)$ theories. This feature has led to a rapidly increasing interest in literature.

Just like in $f(R)$ case, the paper \cite{Gakis:2019rdd} has written down the conformal equivalent scalar-tensor theory of $f(Q)$ case, whose action is
\begin{eqnarray}
	S=\frac{1}{2}\int \sqrt{-\tilde{g}}\left[\tilde{Q}+\sqrt{\frac{2}{3}}(\tilde{\bar{Q}}_{\sigma}{}^{\sigma\mu}-\tilde{Q}^{\mu\sigma}{}_{\sigma})\partial_{\mu}\theta+\tilde{g}^{\mu\nu}\partial_\mu\theta\partial_\nu\theta-2U(\theta)\right] + S_{m}(\tilde{g}_{\mu\nu}/f_s).
\end{eqnarray}
where the quantities which denoted by a tilde over the head are associated the transformed metric $\tilde{g}_{\mu\nu}=f_Qg_{\mu\nu}$ and $\theta=\sqrt{3/2}\ln f_{Q}$. It is easy to see that, unlike in $f(R)$ case, there is a nontrivial scalar non-metricity interaction term $\sqrt{\frac{2}{3}}(\tilde{\bar{Q}}_{\sigma}{}^{\sigma\mu}-\tilde{Q}^{\mu\sigma}{}_{\sigma})\partial_{\mu}\theta$, so the conformal scalar field $\theta$ is non-minimally coupled. This feature can also be found in the torsional teleparallel gravity case presented in \cite{Yang:2010ji}. Meanwhile, since our signature of metric is $\{-1,1,1,1\}$, then the sign of kinetic term of scalar $\phi$ is minus. References \cite{Gakis:2019rdd,Wright:2016ayu,Hu:2023gui} think the scalar $\phi$ is a ghost field and suffer from vacuum instability. But since there is a nontrivial scalar non-metricity coupling, we can not say it is a ghost directly. So in this paper, we study its cosmological perturbation to study this problem.

The outline of the paper is as follows. In Sec. \ref{section2}, we give a brief introduction to $f(Q)$ gravity and write down the corresponding conformal transformed scalar-tensor action. In Sec. \ref{section3}, cosmological applications of the conformal transformed action are studied, containing background equations and linear perturbations. In Sec. \ref{section4}, we show the relations between result in Sec. \ref{section3} and the original $f(Q)$ theory. And we can see that since linear perturbation of $f(Q)$ theory has different degrees of freedom, the theory may suffer from strong coupling problem. Sec. \ref{section5} is a summary of this paper.

\section{f(Q) gravity and conformal transformation}\label{section2}
\subsection{f(Q) gravity}

Symmetric teleparallel gravity (STG) \cite{Nester:1998mp,BeltranJimenez:2017tkd} is a constrained metric-affine theory whose affine connection satisfies the curvature-free and torsion-free conditions
\begin{eqnarray}
	{R_{\mu\nu\sigma}}^{\rho}= - \partial_{\mu} {\Gamma^{\rho}}_{\nu\sigma} + \partial_{\nu} {\Gamma^{\rho}}_{\mu\sigma} +{\Gamma^{\lambda}}_{\mu\sigma}{\Gamma^{\rho}}_{\nu\lambda} - {\Gamma^{\lambda}}_{\nu\sigma}{\Gamma^{\rho}}_{\mu\lambda}=0,\quad {T^{\lambda}}_{\mu\nu} = {\Gamma^{\lambda}}_{\mu \nu} - {\Gamma^{\lambda}}_{\nu\mu}=0,
\end{eqnarray}
then the quantity that describes the gravity effects is non-metricity tensor
\begin{eqnarray}
	Q_{\alpha \mu \nu} =\nabla_{\alpha} g_{\mu \nu} = \partial_{\alpha} g_{\mu\nu} - {\Gamma^{\lambda}}_{\alpha\mu}g_{\lambda\nu} - {\Gamma^{\lambda}}_{\alpha\nu}g_{\mu\lambda},
\end{eqnarray}
where the signature of metric $g_{\mu\nu}$ is $\{-1,1,1,1\}$. The non-zero non-metricity tensor means that the length of vector can change if we parallel transport (described by connection $\Gamma^{\lambda}{}_{\mu\nu}$) it along curves. Since the curvature tensor vanishes, the parallel transport is independent of the path, which is the reason for the terminology – 'teleparallel' and the zero torsion means affine connection is symmetric in its lower indices and which is corresponded to the terminology – 'symmetric'. The differences between STG and GR should be clarified. It can be simply understood as that there are many parallel structures on a manifold, GR chooses one class ($Q=T=0$) of them to describe gravity effects and STG chooses another class ($R=T=0$).

For affine connections satisfy the curvature free and torsion free conditions, we can always choose a particular coordinate system in which $\Gamma^{\lambda}{}_{\mu\nu}=0$. Hence STG theory can be considered as a gravity theory which only takes metric as fundamental variables but does not own diffeomorphism symmetry. There is also a diffeomorphism invariant version of STG. Using St{\"u}eckelberg trick, we can recover the diffeomorphism symmetry. The contents of this tick is as follows. Since there is a particular coordinate system in which affine connection vanishes, in other coordinate systems the connection will become
\begin{eqnarray}\label{2. Gamma=}
	{\Gamma^{\lambda}}_{\mu \nu} = \frac{\partial x^{\lambda}}{\partial y^{\beta}} \partial_{\mu}\partial_{\nu} y^{\beta},
\end{eqnarray}
where $x^{\mu}$ are coordinates and $y^{\mu}$ can be considered as four scalar fields. It can be easily to see that if $y^{\mu}=x^{\mu}$, affine connection becomes zero. Then we can take metric $g_{\mu\nu}$ and $y^{\mu}$ as fundamental variables, where $\{y^{\mu}\}$ represents the particular coordinate system in which affine connection vanishes and now the theory is diffeomorphism invariant.

There is a special STG theory which is called symmetric teleparallel equivalent of general relativity (STEGR) whose Lagrangian density is
\begin{eqnarray}\label{2.nonmetricity scalar}
	Q\equiv -\frac{1}{4}Q_{\alpha\mu\nu}Q^{\alpha\mu\nu} +\frac{1}{2}Q_{\alpha\mu\nu}Q^{\mu\nu\alpha}+\frac{1}{4}Q^{\alpha}Q_{\alpha}-\frac{1}{2}{\bar{Q}}^{\alpha}Q_{\alpha}, 
\end{eqnarray}
where 
\begin{eqnarray}
	Q_{\alpha} = g^{\sigma\lambda}Q_{\alpha\sigma\lambda},\quad \bar{Q}_{\alpha} =  g^{\sigma\lambda}Q_{\sigma\alpha\lambda}
\end{eqnarray}
are two different contractions of non-metricity tensor. It can be proved that 
\begin{eqnarray}
	Q=\mathring{R}+\mathring{\nabla}_{\mu}(Q^{\mu} - \tilde{Q}^{\mu}),
\end{eqnarray}
where $\mathring{R}$ is the scalar curvature corresponding to the Levi-Civita connection which is the connection chosen by GR. Also we should point out that in this paper, all quantities which denoted by a ring over the heads are associated with the
Levi-Civita connection. Therefore the action of STEGR
\begin{eqnarray}\label{2.action Q}
	S_g= \frac{1}{2}\int d^{4}x   \sqrt{-g} Q=\frac{1}{2}\int d^{4}x   \sqrt{-g}\left[  \mathring{R}+\mathring{\nabla}_{\mu}(Q^{\mu} - \tilde{Q}^{\mu}) \right]
\end{eqnarray}
 is only one total derivative away from GR. Since total derivative in action does not influence equations of motion, in this sense, STEGR is equivalent to GR. Because of this equivalence, it is easy to construct modified gravity models with little deviation from GR. $f(Q)$ gravity \cite{Lu:2019hra,BeltranJimenez:2019tme} is a simple case which replaces the non-metricity scalar $Q$ of the action (\ref{2.action Q}) to its arbitrary function $f(Q)$, and this model is what we consider in this paper.

\subsection{Conformal transformation of f(Q) gravity}

The dynamical equivalence between $f(R)$ gravity and a particular scalar-tensor theory is well-studied, both in the case of metric \cite{Futamase:1987ua} and Palatini formalism \cite{Flanagan:2003rb}. Imitating those procedures, paper \cite{Gakis:2019rdd} wrote down the equivalent scalar-tensor theory of $f(Q)$ gravity. Here we briefly repeat the process. 

The conformal transformation of metric can be expressed as
\begin{eqnarray}\label{2.defination of CT}
	\tilde{g}_{\mu\nu}=\Omega^2(x)g_{\mu\nu},\quad  \tilde{g}^{\mu\nu}=\Omega^{-2}(x)g^{\mu\nu},
\end{eqnarray}
which means we will use the new metric $\tilde{g}_{\mu\nu}$ to describe gravity effects. Just as we have said in previous subsection, STG gravity can be considered as a theory of diffeomorphism symmetry breaking and takes the metric as the only fundamental variable. Then under the transformation of Eq.(\refeq{2.defination of CT}), one can find the non-metricity tensor transforms as
\begin{eqnarray}\label{2.Q=tildeQ}
	Q_{\alpha\mu\nu}=\partial_\alpha g_{\mu\nu}=\Omega^{-2}\left(\tilde{Q}_{\alpha\mu\nu}-2\partial_{\alpha}\ln\Omega \,\tilde{g}_{\mu\nu}\right),
\end{eqnarray} 
the non-metricity scalar (\ref{2.nonmetricity scalar}) transforms as
\begin{eqnarray}\label{2. Q CT}
	Q=\Omega^2\left[\tilde{Q}-2(\tilde{Q}^{\mu}-\tilde{\bar{Q}}^{\mu})\partial_\mu\ln\Omega+6\tilde{g}^{\mu\nu}\partial_\mu\ln\Omega\partial_\nu\ln\Omega\right]
\end{eqnarray}
Define boundary term
\begin{eqnarray} \label{2.R=Q+B}
	B=\mathring{\nabla}_{\mu}(\bar{Q}^\mu-Q^{\mu}) \quad\Rightarrow\quad  \mathring{R}=Q+B,
\end{eqnarray}
then it transforms as
\begin{eqnarray}
	B&=&\Omega^{2}\left[\tilde{B}+6\tilde{g}^{\mu\nu}\tilde{\mathring{\nabla}}_\mu\tilde{\mathring{\nabla}}_\nu\ln\Omega+2(\tilde{Q}^{\mu}-\tilde{\bar{Q}}^{\mu})\partial_\mu\ln\Omega-12\tilde{g}^{\mu\nu}\partial_\mu\ln\Omega\partial_\nu\ln\Omega\right].
\end{eqnarray}
It can be easily checked that the scalar curvature corresponding to Levi-Civita connection transforms as usual
\begin{eqnarray}
	R=\Omega^2\left[\tilde{R}+6\tilde{g}^{\mu\nu}\tilde{\mathring{\nabla}}_\mu\tilde{\mathring{\nabla}}_\nu\ln\Omega -6\tilde{g}^{\mu\nu}\partial_{\mu}\ln\Omega\partial_{\nu}\ln\Omega\right].
\end{eqnarray} 

Using Lagrangian multiplier, we can express $f(Q)$ theory as
\begin{eqnarray}\label{2. action fQ}
	S=\int d^4x \sqrt{-g}\left[f(s)+\lambda (Q-s)\right] +S_m(g_{\mu\nu}).
\end{eqnarray}
Variation with respect to $s$ gives
\begin{eqnarray}
	\frac{\partial f(s)}{\partial s}-\lambda=0,
\end{eqnarray}
and taking it back into the action, we obtain
\begin{eqnarray}\label{2. action fQ s}
	S=\int d^4x \sqrt{-g}\left[f_s Q+(f-sf_s)\right] +S_m(g_{\mu\nu}),
\end{eqnarray}
where $f_s=\partial f/\partial s$. It is easy to prove that action (\ref{2. action fQ}) and (\ref{2. action fQ s}) share the same equations of motion. Using Eq.(\ref{2. Q CT}), the action (\ref{2. action fQ s}) can be written as
\begin{eqnarray}
	S=\frac{1}{2}\int d^4x \sqrt{-\tilde{g}}\left[\Omega^{-2}f_s\left(\tilde{Q}-2(\tilde{Q}^{\mu}-\tilde{\bar{Q}}^{\mu})\partial_\mu\ln\Omega+6\tilde{g}^{\mu\nu}\partial_\mu\ln\Omega\partial_\nu\ln\Omega\right) +\Omega^{-4}(f-s f_{s})\right]+S_m(\tilde{g}_{\mu\nu}/\Omega^2).
\end{eqnarray}
Defining the conformal factor $\Omega^2=f_s$ and
\begin{eqnarray}
	\theta=\sqrt{6}\ln\Omega=\sqrt{\frac{3}{2}}\ln f_{s},\quad
	U(\theta)=\frac{1}{2f_s^2}(s f_{s}-f),
\end{eqnarray}
finally we obtain
\begin{eqnarray}\label{2.action Q+tilde}
	S=\frac{1}{2}\int d^4x \sqrt{-\tilde{g}}\left[\tilde{Q}+\sqrt{\frac{2}{3}}(\tilde{\bar{Q}}^{\mu}-\tilde{Q}^{\mu})\partial_{\mu}\theta+\tilde{g}^{\mu\nu}\partial_\mu\theta\partial_\nu\theta-2U(\theta)\right] + S_{m}(\tilde{g}_{\mu\nu}/f_s).
\end{eqnarray}
From here we can find that, not like $f(R)$ theory, $f(Q)$ theory is not equivalent to a theory whose Lagrangian is $Q$ plus a minimaly coupled scalar field, because the second term of action (\ref{2.action Q+tilde}). Meanwhile, since our signature of metric is $\{-1,,1,1,1\}$, the sign of kinetic term of $\theta$ is minus. Just from this action (\ref{2.action Q+tilde}), many papers \cite{Wright:2016ayu,Gomes:2023tur,Hu:2023gui,Gakis:2019rdd}  say it is ghost field, but since the scalar field $\theta$ is also coupled with non-metricity tensor, we can not say it is a ghost directly. Such as a gravity action in Riemann geometry
\begin{eqnarray}
	S=\int d^4x\sqrt{-g} \left[e^{2s}\mathring{R}+cg^{\mu\nu}\partial_\mu s\partial_\nu s\right],
\end{eqnarray}
defining conformal factor $\Omega^2=e^2s$, then the conformal transformed form of this action is
\begin{eqnarray}
	S=\int d^4x\sqrt{-g} \left[\tilde{\mathring{R}}-(6-ce^{2s})\tilde{g}^{\mu\nu}\partial_\mu s\partial_\nu s\right].
\end{eqnarray}
It can be seen that if $6-ce^{2s}>0$ the sign kinetic term of $s$ is plus. It can be easily achieved by requiring a small $c$. Therefore, whether the action (\ref{2.action Q+tilde}) contains ghost fields still needs more studies. 

In this section, we consider metric as the only fundamental variable of STG theory. One can also consider the effects of affine connection. It is easily to show that under conformal transformation, if the affine connection does not change, we can obtain the same result since the relation Eq.(\ref{2.Q=tildeQ})
\begin{eqnarray}
	Q_{\alpha\mu\nu}=\Omega^{-2}\left(\tilde{Q}_{\alpha\mu\nu}-2\partial_{\alpha}\ln\Omega \,\tilde{g}_{\mu\nu}\right),
\end{eqnarray}
is still maintained. One can also consider such transformation under which affine connection changes but still is a STG connection with form (\ref{2. Gamma=}). It is an interesting direction of future studies.

\subsection{Conformal transformation and field redefinition}
Here we summarize what we have done in previous subsection. Using Lagrangian multiplier, we found the equivalent action (\ref{2. action fQ s}) of $f(Q)$ gravity, which is also called the Jordan frame of gravity, and it is the starting point.  Then we did redefinition of fields from $\{g_{\mu\nu}, s\}$ to $\{\tilde{g}_{\mu\nu},\theta\}$ 
\begin{eqnarray}
	\tilde{g}_{\mu\nu}=f_s g_{\mu\nu},\quad \theta=\sqrt{6}\ln\Omega=\sqrt{\frac{3}{2}}\ln f_{s},
\end{eqnarray}
to re-express the action (\ref{2. action fQ s}) to the action (\ref{2.action Q+tilde}) which we will use to analyze the stability. It should be emphasized that the action (\ref{2. action fQ s}) and the action (\ref{2.action Q+tilde}) are just different manifestations of the same action. Hence if the field redefinition is reversible, the equations of motion do not change. This fact is easy to understand. Since the action are the same, then the variation of the action is
\begin{eqnarray}
	\delta S &=& \int d^4x \frac{\delta S}{\delta g_{\mu\nu}}\Big|_s\delta g_{\mu\nu}+\frac{\delta S}{\delta s}\Big|_g\delta s\nonumber\\
	&=&\int d^4x  \frac{\delta S}{\delta g_{\mu\nu}}\Big|_s\left(\frac{\delta g_{\mu\nu}}{\delta \tilde{g}_{\alpha\beta}}\Big|_\theta\delta \tilde{g}_{\alpha\beta}+\frac{\delta g_{\mu\nu}}{\delta \theta}\Big|_{\tilde{g}}\delta \theta\right)+\frac{\delta S}{\delta s}\Big|_g\left(\frac{\delta s}{\delta \theta}\right)\delta\theta,
\end{eqnarray}
then if the field redefinition is reversible, whether the fundamental variables are $\{g_{\mu\nu}, s\}$ or $\{\tilde{g}_{\mu\nu},\theta\}$, it will give the same equations of motion, $\delta S/\delta g_{\mu\nu}|_s=0$ and $\delta S/\delta s|_g=0$. In other words, if the EOM corresponding to $\{g_{\mu\nu}, s\}$ is
\begin{eqnarray}
	H(g,s)=T(g,s),
\end{eqnarray}
then the EOM corresponding to $\{\tilde{g}_{\mu\nu}, \theta\}$ is
\begin{eqnarray}
	H(g(\tilde{g},\theta),s(\theta))=T(g(\tilde{g},\theta),s(\theta)).
\end{eqnarray}
Hence for $f(Q)$ case, if we find a solution $\{\tilde{g}, \psi\}$ for the EOMs of conformal transformed action (\ref{2.action Q+tilde}), then $\{\tilde{g}/f_s, s\}$ is a  solution of the EOMs of action (\ref{2. action fQ s}). There are many works to do this solution transformation.

In this paper, the conformal transformation is just a mathematical tool to simplify our calculation and from other point of view to check the stability of $f(Q)$ theory. There are also many papers \cite{Faraoni:1998qx,Levy:2015awa} to talk about the physical meanings of conformal transformation.

\section{Cosmological application}\label{section3}

\subsection{Action and equations of motion}

The main purpose of this paper is to study original action by using conformal transformed action. For $f(Q)$ gravity, the conformal transformed action is Eq.(\ref{2.action Q+tilde}), and we will study its cosmological applications in this section. For writing convenient, we will omit the tilde over the transformed quantities in this section, hence the action we study here is
\begin{eqnarray}\label{3.action Q+theta}
	S=\frac{1}{2}\int \sqrt{-g}\left[Q+\sqrt{\frac{2}{3}}(\bar{Q}^{\mu}-Q^{\mu})\partial_{\mu}\theta+g^{\mu\nu}\partial_\mu\theta\partial_\nu\theta-2U(\theta)\right] + S_{m}(g_{\mu\nu}/f_s).
\end{eqnarray}
The arbitrariness of the functional form of $f(Q)$ is reflected in the form of potential function $U(\theta)$. The action of matter sector we choose is
\begin{eqnarray}
	S_{m}=\frac{1}{2}\int d^4x\sqrt{-g}P(Y,\phi),
\end{eqnarray}
where
\begin{eqnarray}
	Y=f_sg^{\mu\nu}\partial_{\mu}\phi\partial_{\nu}\phi,\quad f_s = e^{\sqrt{\frac{2}{3}}\theta},
\end{eqnarray}
and it is just transformed k-essence action. The fundamental variables are scalar fields $\phi$, $\theta$, metric $g_{\mu\nu}$ and constraint affine connection ${\Gamma^{\lambda}}_{\mu \nu} = \frac{\partial x^{\lambda}}{\partial y^{\beta}} \partial_{\mu}\partial_{\nu} y^{\beta}$. Because the constraint affine connection can be considered as the St{\"u}eckelberg fields to restore the broken diffeomorphism symmetry, it can be shown that the equation of motion corresponding to affine connection is contained by the EOM of metric \cite{Zhao:2021zab,Li:2021mdp}. Hence here we will do not write the EOM of connection.

The variation of action (\ref{3.action Q+theta}) with respect to metric gives the metric EOM
\begin{eqnarray}
	\mathring{G}_{\mu\nu}+\sqrt{\frac{2}{3}}\left(\frac{1}{2}\bar{Q}^\lambda\partial_{\lambda}\theta g_{\mu\nu}-\frac{1}{2}\partial_{(\mu}\theta Q_{\nu)}+\nabla_{\mu}\nabla_{\nu}\theta-\nabla^{\lambda}\nabla_{\lambda}\theta g_{\mu\nu}\right)=T^{\theta}_{\mu\nu}+T^{\phi}_{\mu\nu},
\end{eqnarray}
where $\mathring{G}_{\mu\nu} = \mathring{R}_{\mu\nu}-1/2\mathring{R}g_{\mu\nu}$ is the Einstein tensor corresponding to Levi-Civita connection and
\begin{eqnarray}
	T^{\theta}_{\mu\nu}=-\partial_{\mu}\theta\partial_{\nu}\theta+ \frac{1}{2}g_{\mu\nu}(g^{\alpha\beta}\partial_{\alpha}\theta\partial_{\beta}\theta-2U),\\
	T^{\phi}_{\mu\nu}=-P_{Y}f_{s}\partial_{\mu}\phi\partial_{\nu}\phi+\frac{1}{2}Pg_{\mu\nu},
\end{eqnarray}
are energy-momentum tensor of $\theta$ and $\phi$ separately. Meanwhile we can get the EOM of $\theta$
\begin{eqnarray}
	-\mathring{\nabla}^{\mu}\mathring{\nabla}_{\mu}\theta-U_{\theta}+\frac{1}{2}P_YY\frac{f_{ss}}{f_s}s_{\theta}-\sqrt{\frac{1}{6}}\mathring{\nabla}_{\mu}\left(\bar{Q}^{\mu}-Q^{\mu}\right)=0,
\end{eqnarray}
and the EOM of $\phi$ is
\begin{eqnarray}
	-2\mathring{\nabla}_{\mu}\left(P_Yf_s\partial^{\mu}\phi\right)+P_{\phi}=0.
\end{eqnarray}
Using those EOMs, we can find the background solutions with any spacetime symmetries.

For cosmological background, the metric form in Cartesian coordinate system is 
\begin{eqnarray}\label{3. metric cartesian}
	ds^2=a(\eta)^2(-d\eta^2+dx^2+dy^2+dz^2).
\end{eqnarray}
Since the curvature free and torsion free condition means that there is a particular coordinate system in which $\Gamma^{\lambda}{}_{\mu\nu}=0$, here we just assume that  Cartesian coordinate system is that particular coordinate system (there are other choices of connection which we will talk about in next section). This assumption has been chosen in many papers and give the consistent results. Then taking metric (\ref{3. metric cartesian}) and $\Gamma^{\lambda}{}_{\mu\nu}=0$ into the equations of motion, we obtain the equations of cosmological background
\begin{eqnarray}
	3\mathcal{H}^2-\sqrt{6}\mathcal{H}\theta'=-\frac{1}{2}\theta^{'2}+a^2U-P_1\phi^{'2}-\frac{1}{2}a^2P\label{3.eom metric},\\
	\mathcal{H}^2-2\frac{a''}{a}-\sqrt{\frac{2}{3}}\mathcal{H}\theta'+\sqrt{\frac{2}{3}}\theta''=-\frac{1}{2}\theta^{'2}-a^2U+\frac{1}{2}a^2P,\\
	\theta''+2\mathcal{H}\theta'-a^2U_{\theta}-\sqrt{6}\left(\mathcal{H}^2+\frac{a''}{a}\right)+H(\theta,\phi)=0,\\
	P_1\left(\phi''+2\mathcal{H}\phi'\right)+\frac{1}{2}a^2P_{\phi}+P'_{1}\phi'=0,
\end{eqnarray}
where the first two equations are metric equations and we have define
\begin{eqnarray}
	P_1=P_{Y}f_s,\quad H(\theta,\phi)=\frac{a^2}{2}P_YY\frac{f_{ss}}{f_s}s_{\theta}.
\end{eqnarray}
Finding solutions of those equations, we can get the evolution of universe. Relevant contents can be referred to \cite{Hama:2023bkl}. In the following, we will use those equations to get the quadratic action of perturbation variables.

\subsection{Cosmological perturbations}

Now we turn to the linear perturbations around this background. The line element of perturbed metric is
\begin{eqnarray}
	ds^2=a(t)^2\left[-(1+2A)d\eta^2-2(\partial_iB+B_i)d\eta dx^i+\left((1-2\psi)\delta_{ij}+2\partial_i\partial_j E+\partial_iE_j+\partial_jE_i+h_{ij}\right)dx^idx^j\right],
\end{eqnarray} 
where we have used scalar-vector-tensor (SVT) decomposition. $A$, $B$, $\psi$, $E$ are four scalar perturbations, $B_i$, $E_i$ are four vector perturbations satisfying transverse condition, $\partial_iB_i=\partial_iE_i=0$, and $h_{ij}$ are two tensor perturbations satisfying transverse and traceless conditions, $\partial_ih_{ij}=\delta^{ij}h_{ij}=0$. For the constraint affine connection, since its background value is zero, we can choose the fundamental variables $y^{\mu}$ of Eq.(\ref{2. Gamma=}) as $y^{\mu}=x^{\mu}+u^{\mu}$, where $x^{\mu}$ are background, $u^{\mu}$ are perturbations. Imitating SVT decomposition of metric, we can decompose $u^{\mu}$ as 
\begin{eqnarray}
	u^{\mu}=\{u^0,\partial_iu+u_i\}.
\end{eqnarray}
Gauge transformation play important roles in cosmological perturbation theory. Since the gravity action is diffeomorphism invariant, linear perturbation equations can be written only using gauge invariant variables. Then under a gauge transformation $x^\mu \rightarrow x^{\mu}+\zeta^{\mu}$, where $\zeta^{\mu}=\{\zeta^0, \partial_i\zeta+\zeta_i\}$, the perturbation variables $u^{\mu}$ transform as
\begin{eqnarray}
	u^0\rightarrow u^0-\zeta^0,\quad u\rightarrow u-\zeta,\quad u_i\rightarrow u_i-\zeta_i.
\end{eqnarray}
From this transformation rule, one can choose the so-called coincident gauge, $u^0=u=u_i=0$, to calculate perturbation equations. This gauge is used in almost all the papers which calculate perturbations in the framework of STG theory. But in this paper, we will use the so-called unitary gauge
\begin{eqnarray}
	E=E_i=0, \quad \delta\theta=0
\end{eqnarray}
 which are frequently used in GR framework to calculate. One reason to choose unitary gauge is that we can easily compare the results with GR, the other is that we can see the behaviors of St{\"u}eckelberg field $u^{\mu}$ more clearly.

We will calculate the quadratic action for perturbations in the following. From this we can not only obtain the linear perturbation equations, but also analyze if there are ghost fields in theory. Meanwhile since in the linear cosmological perturbation theory, the scalar, vector and tensor perturbations evolve independently, we will deal with them separately.

The quadratic action of tensor perturbations is
\begin{eqnarray}
	S^{(2)}_{T}=\int d^4x a^2\left\{\frac{1}{8}(h'_{ij}h'_{ij}-\partial_{l}h_{ij}\partial_{l}h_{ij})-(2\mathcal{H}^2+4\mathcal{H}')h_{ij}h_{ij}+\left[\frac{1}{2\sqrt{6}}(\theta''-2\theta' \mathcal{H})+\frac{1}{8}(\theta^{'2}+2a^2U)-\frac{1}{8}a^2P\right]h_{ij}h_{ij}\right\}\nonumber\\
\end{eqnarray}
and taking the background equation (\ref{3.eom metric}) of metric back into this action, we get the final result
\begin{eqnarray}
	S^{(2)}_{T}=\int d^4x\frac{a^2}{8}\left(h'_{ij}h'_{ij}-\partial_{l}h_{ij}\partial_{l}h_{ij}\right),
\end{eqnarray}
which is the same as GR. This means that gravitational waves (GWs) behavior of action (\ref{3.action Q+theta}) are the same as GR. But since the action (\ref{3.action Q+theta}) is the conformal transformed action of $f(Q)$, this dose not mean that the $f(Q)$ gravity does not change the GWs. We leave this topic for the next section.

The quadratic action of vector perturbations is
\begin{eqnarray}
	S^{(2)}_{V}=\int d^4x \frac{1}{4}a^2\left[\partial_iB_j\partial_iB_j+\left(6\mathcal{H}^2-2\sqrt{6}\mathcal{H}\theta'+\theta^{'2}-2a^2U+a^2P+2P_1\phi^{'2}\right)B_iB_i\right]
\end{eqnarray}
and taking the background equation (\ref{3.eom metric}) of metric back into this action, we get the final result
\begin{eqnarray}
	S^{(2)}_{V}=\int d^4x \frac{1}{4}a^2\partial_iB_j\partial_iB_j,
\end{eqnarray}
which is also the same as GR. There are no propagating modes of vector perturbations.

The quadratic action of scalar perturbations is more complicated. The gravitational component is
\begin{eqnarray} \label{3.quadratic action scalar}
	S^{(2)}_{S}=\int d^4x a^2\left[\Delta B(2\psi'+2\mathcal{H}A-\frac{3}{2}\mathcal{H}^2B)-9\mathcal{H}^2A\psi-6\mathcal{H}A\psi'+2A\Delta\psi-\frac{9}{2}\mathcal{H}^2A^2+\frac{3}{2}\mathcal{H}^2\psi^2+3\mathcal{H}'\psi^2-\psi\Delta\psi-3\psi^{'2}\right]\nonumber\\
	+a^2\frac{\theta'}{\sqrt{6}}\left[\Delta B(\Delta u-\partial_tu^0-A+3\psi+3\mathcal{H}B)+A(-\Delta u^0+\partial_t\Delta u+6\partial_t\psi+9\mathcal{H}A+18\mathcal{H}\psi)\right.\nonumber\\\left.+\psi\Delta u^0+3\psi\partial_t\Delta u+6\psi\partial_t\psi+9\mathcal{H}\psi^2+\partial_tu^0\Delta u^0-\Delta u\Delta u^0+\partial_t\Delta u\Delta u-\partial_t u^0\partial_t \Delta u\right]\nonumber\\
	+a^2\left[(-\frac{3}{4}\theta^{'2}+\frac{1}{2}a^2U)A^2+(-\frac{3}{2}\theta^{'2}+3a^2U)A\psi+(-\frac{3}{4}\theta^{'2}-\frac{3}{2}a^2U)\psi^2+(-\frac{1}{4}\theta^{'2}+\frac{1}{2}a^2U)B\Delta B\right],\nonumber\\
\end{eqnarray}
and the matter component is
\begin{eqnarray}
	S^{(2)}_m=\int d^4x -\frac{1}{4}a^4P(A^2+6A\psi-3\psi^2-\partial_iB\partial_iB)+\frac{1}{2}a^4P_\phi(A\delta\phi-3\psi\delta\phi)\nonumber\\
	+a^2P_YF(-\frac{1}{2}\partial_t\delta\phi\partial_t\delta\phi+\frac{1}{2}\partial_i\delta\phi\partial_i\delta\phi+\phi'A\partial_t\delta\phi-\phi^{'2}A^2+3\phi'\psi\partial_t\delta\phi-3\phi^{'2}A\psi-\phi'\partial_iB\partial_i\delta\phi+\frac{1}{2}\phi^{'2}\partial_iB\partial_iB)\nonumber\\
	+\frac{1}{4}a^4P_{\phi\phi}\delta\phi^2+P_{YY}F^2(\phi^{'2}\partial_t\delta\phi\partial_t\delta\phi-2\phi^{'2}A\partial_t\delta\phi+\phi^{'4}A^2)+a^2P_{Y\phi}F(-\phi'\delta\phi\partial_t\delta\phi+\phi'^2A\delta\phi).\nonumber\\
\end{eqnarray}
One can see that the variable $B$ is not dynamical field, variation with respect to it gives
the constraint equation,
\begin{eqnarray} \label{3.constraint B}
	\Delta B\left[2\psi'+2\mathcal{H}A+\frac{\theta'}{\sqrt{6}}(\Delta u-\partial_tu^0-A+3\psi)+P_1\phi'\delta\phi\right]=0.
\end{eqnarray} 
This constraint equation can be considered as an equation to get the relation of field $A$
\begin{eqnarray}\label{3. A relation}
	A=2M\psi'+\frac{M\theta'}{\sqrt{6}}(\Delta u-\partial_tu^0+3\psi)+MP_1\phi'\delta\phi,
\end{eqnarray}
where we have define $M=1/(\theta'/\sqrt{6}-2\mathcal{H})$. We can easily find that there is also no term containing time derivative of $A$ in quadratic action, then perturbation $A$ is also an auxiliary field. Therefore taking the relation (\ref{3. A relation}) back into the quadratic action of scalar perturbations, we will get another quadratic action only relies on variables $\{\psi, u^0, u, \delta\phi\}$. Since the perturbation $A$ does not relate to the term $\partial_t\delta\phi$, the kinetic term of perturbation $\delta\phi$ is
\begin{eqnarray}
	\int d^4x \left(-\frac{1}{2}a^2P_YF+P_{YY}F^2\phi^{'2}\right)\partial_t\delta\phi\partial_t\delta\phi,
\end{eqnarray}
which is no mix term with perturbations $\{\psi, u^0, u\}$. We can see that if $F=1$ and $P=-Y$ which corresponds to a minimally coupled scalar field, the kinetic term of $\phi$ is proportional to $a^2/2$ which is always positive-defined. Since there is no mix term of kinetic term of $\delta\phi$ with other perturbations, we will omit the matter component and only consider the gravity part.

When the matter sector is empty, the equation of perturbation A gives another constraint
\begin{eqnarray}
	2\mathcal{H}\Delta B-9\mathcal{H}^2\psi-6\mathcal{H}\psi'+2\Delta \psi +\frac{\theta'}{\sqrt{6}}(-\Delta B-\Delta u^0+\partial_t\Delta u+\partial_t\psi+18\mathcal{H}\psi)
	+(-\frac{3}{2}\theta^{'2}+3a^2U)\psi-2a^2UA=0,
\end{eqnarray}
and this constraint can be considered as an equation solving $\Delta B$. From Eq.(\ref{3.constraint B}) we knew that $\Delta B$ is something like a Lagrangian multiplier, so this constraint does not influence the final form of quadratic action of scalar perturbations. Using the relation (\ref{3. A relation}), we can write down the related kinetic term of quadratic action
\begin{eqnarray}
	S^{(2)}_{S\,kin}=\int d^4x M^2\left[(-\frac{1}{2}\theta^{'2})\psi'\psi'+(-\mathcal{H}\theta'+\frac{1}{\sqrt{6}}\theta^{'2})\theta'\psi'u^{0'}+(-2\sqrt{\frac{2}{3}}\mathcal{H}\theta'+\frac{1}{3}\theta^{'2})\psi\Delta u'\right.\nonumber\\
	\left.+(-\frac{1}{2}\mathcal{H}^2+\frac{1}{\sqrt{6}}\mathcal{H}\theta'-\frac{1}{12}\theta^{'2})\theta^{'2}u^{0'}u^{0'}+(-2\frac{2}{3}\mathcal{H}^2+\mathcal{H}\theta'-\frac{1}{3\sqrt{6}}\theta^{'2})\theta'u^{0'}\Delta u'
	\right],
\end{eqnarray}
and this action can be written as a more compact form
\begin{eqnarray}
	S^{(2)}_{S\,kin}=\int d^4x M^2 \Phi^{T}K\Phi,
\end{eqnarray}
where $\Phi=\{\psi', \theta'u^{0'},\Delta u'\}$ and $K$ is kinetic matrix. It is easy to show that this kinetic matrix has three eigenvalues
\begin{eqnarray}
	&&\lambda_0=0,\nonumber\\ &&\lambda_{1,2}=\frac{1}{72}(-18\mathcal{H}^2+6\sqrt{6}\mathcal{H}\theta'-21\theta^{'2})\pm \sqrt{3}\sqrt{1260\mathcal{H}^4-648\sqrt{6}\mathcal{H}^2\theta'+2100\mathcal{H}^2\theta^{'2}-324\sqrt{6}\mathcal{H}\theta^{'3}+203\theta^{'4}}.
\end{eqnarray}
The eigenvector corresponds to $0$ is 
\begin{eqnarray}
	\{\frac{\theta'}{\sqrt{6}}-\mathcal{H},1,0\}
\end{eqnarray}
which means we can define three new variables
\begin{eqnarray}
	\alpha_0=(\frac{\theta}{\sqrt{6}}-\ln a)\psi+\theta' u^{0},\quad \alpha_2=\psi -(\frac{\theta}{\sqrt{6}}-\ln a) \theta' u^{0},\quad \alpha_2=\Delta u,
\end{eqnarray}
and $\alpha_0$ does not have kinetic term. If we follow the usual process, the next we should do is to use those new variables $\alpha$ to rewrite quadratic action of scalar perturbations (\ref{3.quadratic action scalar}) and obtain the constraint of $\alpha_0$. But from Refs.\cite{Gomes:2023tur,Rao:2023nip} we know that $\alpha_0$ is a dynamical variable on other background. At linear perturbation level, if a model shows different degrees of freedom on different backgrounds, it means linear perturbation theory breaks down on the background of decreasing number of degrees of freedom and may suffer from strong coupling issue. Since the linear perturbation is not reliable, we need analyze its higher order behaviors.

\section{Transformation rules and strong coupling problem}\label{section4}

\subsection{The transformation rules of perturbation variables}
In the previous section, we studied the cosmological applications of conformal transformed action (\ref{3.action Q+theta}) of $f(Q)$ model. Particularly, we calculated the linear cosmological perturbation of the action (\ref{3.action Q+theta}). Then in this part, we will talk about how to relate the transformed result and the original model.

The perturbed cosmological metric of original model can be written as
\begin{eqnarray}
	ds^2=a(t)^2\left[-(1+2A)d\eta^2-2(\partial_iB+B_i)d\eta dx^i+\left((1-2\psi)\delta_{ij}+2\partial_i\partial_j E+\partial_iE_j+\partial_jE_i+h_{ij}\right)dx^idx^j\right].
\end{eqnarray} 
Perturbed conformal factor can be expanded as
\begin{eqnarray}
	f_s\equiv F(s), \quad  F(s)=F_0+\frac{\partial F}{\partial s}\delta s,
\end{eqnarray}
and since the background of transformed metric $\tilde{g}_{\mu\nu}=F(s)g_{\mu\nu}$ also should be homogeneous and isotropic, its perturbed form can be also written as
\begin{eqnarray}
	d\tilde{s}^2&=&Fds^2\nonumber\\
	&=&\tilde{a}(t)^2\left[-(1+2\tilde{A})d\eta^2-2(\partial_i\tilde{B}+\tilde{B}_i)d\eta dx^i+\left((1-2\tilde{\psi})\delta_{ij}+2\partial_i\partial_j \tilde{E}+\partial_i\tilde{E}_j+\partial_j\tilde{E}_i+\tilde{h}_{ij}\right)dx^idx^j\right],
\end{eqnarray}
Finally we have the relations of perturbation variables
\begin{eqnarray}
	\tilde{A}=A+\frac{1}{2}\frac{\partial\ln F_0}{\partial s}\delta s,\nonumber\\
	\tilde{\psi}=\psi-\frac{1}{2}\frac{\partial\ln F_0}{\partial s}\delta s,\nonumber
\end{eqnarray}
and other perturbation variables are not changed. Background quantities changes as follows:
\begin{eqnarray}
	\tilde{a}^2=F_0a^2,\quad dt=a d\eta = \frac{1}{\sqrt{F_0}}\tilde{a}d\eta=\frac{1}{\sqrt{F_0}}d \tilde{t},
\end{eqnarray}
and
\begin{eqnarray}
	\tilde{\mathcal{H}}=\mathcal{H}+\frac{d\ln\sqrt{F_0}}{d\eta},\quad \tilde{H}=\frac{1}{\sqrt{F_0}}\left[H+\frac{d\ln\sqrt{F_0}}{dt}\right].
\end{eqnarray}
From these transformation rules, we can find that the transformed perturbation variables are just recombinations of the ordinary variables. Since we also need to recombine the variables to diagonalize kinetic matrix to analyze whether there are ghost, it means that the conformal transformation or other non-singular transformations \cite{Bettoni:2013diz,Takahashi:2023vva}, such as non-singular disformal transformation, does  not change the ghost behavior. This is also the reason we can use the transformed action to study the original action. Meanwhile We can define the gauge invariant Mukhanov-Sasaki variable
\begin{eqnarray}
	\zeta=\phi+\mathcal{H}\frac{\delta s}{s'},
\end{eqnarray}
then we have relation
\begin{eqnarray}
	\tilde{\zeta}&=&\tilde{\phi}+\tilde{\mathcal{H}}\frac{\delta s}{s'}\nonumber\\
	&=&\phi-\frac{1}{2}\frac{\partial\ln F_0}{\partial s}\delta s+(\mathcal{H}+\frac{d\ln F_0^{1/2}}{d\eta})\frac{\delta s}{s'}\nonumber\\
	&=&\zeta,
\end{eqnarray}
 which means this variable is a conformal invariant variable \cite{Koh:2005qp,Gong:2011qe}. 

Therefore, using the transformation rules given above, we can write down the final quadratic action of perturbations of $f(Q)$ theory. For background quantities, we have 
\begin{eqnarray}
	\tilde{a}^2\rightarrow f_{Q}a^2, \quad \theta\rightarrow \sqrt{\frac{3}{2}}\ln f_{Q}.
\end{eqnarray}
Then the quadratic action for tensor perturbations of $f(Q)$ model is
\begin{eqnarray}
	S^{(2)}_{T}=\int d^4x\frac{f_{Q}a^2}{8}\left(h'_{ij}h'_{ij}-\partial_{l}h_{ij}\partial_{l}h_{ij}\right),
\end{eqnarray}
and we need to require $f_Q>0$ to avoid ghost modes. Meanwhile, since conformal transformation is $\tilde{g}_{\mu\nu}=f_Q g_{\mu\nu}$, $f_Q>0$ also means that the transformed metric shares the same signature with the original metric. From this quadratic action we can get the equations of gravitational waves (GWs)
\begin{eqnarray}
	h''_{ij}+\left(\frac{f'_{Q}}{f_{Q}}+2\mathcal{H}\right)h'_{ij}+\partial_l\partial_l h_{ij}=0.
\end{eqnarray}
Then we know that the dissipative terms of this wave equation in $f(Q)$ model are different compared to GR, and it will influence the amplitude of gravitational waves. Therefore by detecting the amplitude of the gravitational waves we can limit the parameters in the model. The quadratic action of vector perturbations is
\begin{eqnarray}
	S^{(2)}_{V}=\int d^4x \frac{1}{4}f_{Q}a^2\partial_iB_j\partial_iB_j.
\end{eqnarray}
Since there are no kinetic term of vector perturbations, the vector modes do not propagate, just like them in GR. The quadratic action of scalar perturbations are complicated, we are not going to write it here. But it should be clarified that since the conformal transformation leads to just variables recombination, there are also only $2$ scalar gravitational degrees of freedom show up on the background ($\Gamma^{\lambda}{}_{\mu\nu}=0$) in $f(Q)$ gravity. But from papers \cite{Gomes:2023tur,Rao:2023nip} we know the disappeared variable $\alpha_0$ shows up on other background. At linear perturbation level, some variables vanished on some particular backgrounds always means that the linear perturbation theory breaks down and suffer from strong coupling problem. So we stop our calculation here and talk about related questions in the following subsection.

\subsection{Strong coupling problem}

As we have said in Sec. \ref{section2}, STG theory is a constrained metric-affine theory, and it takes metric and constrained affine connection (\ref{2. Gamma=}) as its fundamental variables. According to the symmetries of flat space FRW universe, we chose the form of metric as
\begin{eqnarray}
	ds^2=-a^2(d\eta^2+dx^2+dy^2+dz^2).
\end{eqnarray}
But how to choose suitable constrained affine connection still need more consideration. In Sec. \ref{section3}, we chose $\Gamma^{\lambda}{}_{\mu\nu}=0$ for simplicity. There are also other choices, such as requiring the affine connection also to satisfy the symmetries of spacetime \cite{Hohmann:2019fvf,Hohmann:2019nat,Hohmann:2021ast}. This requirement can be expressed as $\mathcal{L}_{\zeta}\Gamma^{\lambda}{}_{\mu\nu}=0$, where $\mathcal{L}$ is Lie derivative and $\zeta$ are all Killing vectors of background. Following this requirement, there are three branches of the form of affine connection. Here we only write down the relevant third branch in \cite{Rao:2023nip}, and the non-zero components of affine connection is
\begin{eqnarray}\label{4.gamma}
	\Gamma^0{}_{00}=K_1(\eta),\quad \Gamma^0{}_{i j}=K_2(\eta) \delta_{i j},
\end{eqnarray}
with $K_1=-\gamma'/\gamma$, $K_2=\gamma$. Where $\gamma=\gamma(\eta)$ is an arbitrary function and 'prime' represent the derivative with respect to the conformal time $\eta$. It should be clarified that different forms of affine connection mean different backgrounds. And on this different background, the papers \cite{Gomes:2023tur,Rao:2023nip} have shown that there are $7$ degrees of freedom at linear perturbation level, i.e., two tensor modes, two vector modes and three scalar modes. But we have shown that only two scalar modes and two tensor modes appear on $\Gamma^{\lambda}{}_{\mu\nu}=0$ background.

It is easy to understand that a model shows different degrees of freedom on different backgrounds, since the backgrounds we usually study has a high degree of symmetry . Such as if the kinetic term of a perturbation variable $\lambda$ is $\partial_x \bar{\phi} \times\dot{\lambda}^2(t,x,y,z)$, where $\bar{\phi}$ is background field, we can find that on cosmological case the background field $\bar{\phi}=\bar{\phi}(t)$, there is no kinetic term of $\lambda$. For $f(Q)$ gravity on Minkowski background, the Lagrangian can be expanded perturbatively as
\begin{eqnarray}
	f(Q)=f(Q_0)+f_Q\delta Q+\frac{1}{2}f_{QQ}\delta Q\delta Q+...,
\end{eqnarray}
and since scalar $Q$ is quadratic in non-metricity tensor $Q_{\alpha\mu\nu}$, hence at linear perturbation level  only $f_Q\delta Q$ term survives. Meanwhile we know that scalar $Q=\mathring{R}+B$ from Eq.(\ref{2.R=Q+B}), where $B$ is boundary term, so the linear perturbation of $f(Q)$ gravity is the same as GR. It means that only two tensor modes appear on Minkowski background. This phenomenon appears in many modified gravity models, such as nonlinear extension of Fierz-Pauli action in massive gravity case \cite{Hinterbichler:2011tt,deRham:2014zqa} and $f(Q)$ case \cite{Hu:2023juh,Hu:2023xcf}. This phenomenon is always linked to the strong coupling problem \cite{Charmousis:2009tc}. 

For $f(Q)$ gravity case, the papers \cite{Gomes:2023tur,Rao:2023nip} have shown that the on the background (\ref{4.gamma}) quadratic action of vector perturbations in momentum space
\begin{eqnarray}\label{4.vector quadratic action}
	S_V^{(2)}=\int d\eta d^3k a^2N^2\left[u_i^{'2}-\omega^2u_i^2\right],
\end{eqnarray}
where
\begin{eqnarray}
	N^2=\frac{K_2f'_Q}{1+2k^{-2}K_2(\ln f_Q)'},\quad \omega^2=k^2+2K_2(\ln f_Q)'.
\end{eqnarray}
It is easy to find that if $K_2=0$ which means $\Gamma^{\lambda}{}_{\mu\nu}=0$, there is no kinetic term of vector perturbations, just like we have shown in Sec. \ref{section3}. Here we study the behaviors as $K_2\rightarrow0$. It is not a good idea to take this limit in the form of Eq.(\ref{4.vector quadratic action}). So we define new variables $v_i=Nu_i$, then the quadratic action (\ref{4.vector quadratic action}) becomes
\begin{eqnarray}\label{4.new form of vq}
	S_V^{(2)}=\int d\eta d^3k a^2\left[v_i^{'2}-\omega^2v_i^2+\frac{2\mathcal{H}N'+N''}{N}v_i^2\right].
\end{eqnarray}
The third term of Eq.(\ref{4.new form of vq}) is a mass term of vector perturbations, and as $K_2\rightarrow0$, if $K'_2\neq 0$, this mass term goes to infinity. A particle with large mass are not excited at low energy scale. But for scalar perturbation case, the situation is different. It has been shown that there are three gravitational scalar degrees of freedom, and the determinant of kinetic matrix is
\begin{eqnarray}
	\frac{36k^4K_2^2f'_Qf_Q^2f_{QQ}}{2a^2K_2f'_{Q}+k^2f_{QQ}(2\mathcal{H}-K_1)^2}.
\end{eqnarray}
When $K_2\rightarrow 0$, the determinant of kinetic matrix is zero, which means that there are only two scalar degrees of freedom just like we have shown in Sec. \ref{section3}. But since the scalar perturbations are coupled with each other, the $f(Q)$ theory may suffer from strong coupling problem at linear perturbation level. This situation can be understood as follows. For a simple quadratic action of perturbation variables $x_1$ and $x_2$ 
\begin{eqnarray}
	S=\int d\eta\, a_1(\eta)x_{1}^{'2}+a_2(\eta)x_{2}^{'2}+a_3(\eta)x_1x_2+a_4(\eta)x^2_1,
\end{eqnarray}
where $a_1$, $a_2$, $a_3$ and $a_4$ are background quantities. If $a_1=0$, there are only one degrees of freedom, i.e, $x_2$. If we want to study the behaviors as $a_1(\eta)\rightarrow 0$, we can define new variable $y=a_1x_1$, then the action becomes
\begin{eqnarray}
		S=\int d\eta\, y^{'2}+a_2(\eta)x_{2}^{'2}+\frac{a_3(\eta)}{a_1(\eta)}yx_2+\frac{a_4(\eta)+a_1(\eta)a_1''(\eta)}{a_1^2(\eta)}y^2.
\end{eqnarray}
One can see that the coupling coefficient of $y$ and $x_2$ goes to infinity as $a_1(\eta)\rightarrow 0$. Which means variables $y$ and $x_2$ are strongly coupled and linear perturbation theory breaks down in this situation.

STG theory can be understood as a theory without diffeomorphism symmetry which takes metric as fundamental variable. So there are often more degrees of freedom than GR. In appendix \ref{appendix}, one can find the ADM form of scalar $Q$. From that form, we know that scalar $Q$ contains the time derivatives of lapse function and shift vector, which corresponding to diffeomorphism symmetry in GR case. Therefore it is not surprising that the $f(Q)$ theory has more than two degrees of freedom. Other STG models have similar properties. So it is common that a STG model shows different degrees of freedom on different backgrounds at linear perturbation level, which makes the linear perturbation theory breaks down. There are two ways to deal with this situation. One is to analyze the higher-order perturbation effects of the model \cite{Hu:2023juh,Hu:2023xcf}, another is to add more gravitational interaction terms to make all variables show up at linear order \cite{Zhang:2024vfw,Li:2021mdp}.

\section{Conclusion} \label{section5}

In this paper, we studied the conformal equivalent scalar-tensor theory of $f(Q)$ theory. Although the sign of kinetic term of scalar is minus, we explained that because the scalar is non-minimally coupled with gravity, whether it leads to ghost instability still needs to be studied. Then we studied cosmological applications of the conformal transformed action, especially the linear perturbation theory. We showed that on a particular background, there are two tensor and two scalar propagating modes. Considering the results in the papers \cite{Gomes:2023tur,Rao:2023nip} that $f(Q)$ gravity has $7$ degrees of freedom and on other cosmological background all those $7$ degrees of freedom appear, it means that $f(Q)$ theory shows different degrees of freedom on different backgrounds at linear perturbation level. We explained that this situation always means that the linear perturbation theory breaks down and the $f(Q)$ theory may suffer from strong coupling problem. Therefore, a higher-order perturbation analysis is required for $f(Q)$ gravity. We can also add more gravitational interaction terms to make all variables show up at linear order. We think our study clearly show that it is necessary to know the degrees of freedom of modified gravity models before calculating the perturbation to avoid the strong coupling problem.

\subsection*{Acknowledgements}
We would like to thank Yunsong Piao, Jun Zhang and Yeheng Tong for very useful discussion. This work is supported by NSFC, No.12075246 and the scientific research starting grants from University of Chinese Academy of Sciences (grant No.118900M152).

\appendix

\section{3+1 decomposition of STG gravity}\label{appendix}

 STG can be considered as a theory which takes metric as the only fundamental variable without diffeomorphism invariant. Hence lapse function and shift vector are usually dynamical fields in STG theory. Here we will study its 3+1 spacetime decomposition. Metric can be written as
\begin{eqnarray}
	ds^2=-(N^2-N^iN_i)dt^2+2N_idtdx^i+h_{ij}dx^idx^j,
\end{eqnarray}
where $N$ is lapse function, $N^i$ is shify vector. Choosing coincident gauge, non-metricity tensor is
\begin{eqnarray}
	Q_{\alpha\mu\nu}=\partial_\alpha g_{\mu\nu}.
\end{eqnarray}
Define projective tensor
\begin{eqnarray}
	h_{\mu\nu}=g_{\mu\nu}+n_\mu n_\nu
\end{eqnarray}
where the components of normal vector $n$ are
\begin{eqnarray}
	n_{\mu}=\{-N,0,0,0\},\quad n^\mu=\{1/N, -N^i/N\}.
\end{eqnarray}
We can also define extrinsic Curvature tensor, whose spatial components are
\begin{eqnarray}
	K_{ij}=\frac{1}{2N}(h'_{ij}-D_iN_j-D_jN_i)=\frac{1}{2N}(h'_{ij}-h_{im}\partial_jN^m-h_{jm}\partial_iN^m-N^m\partial_m h_{ij})\\
	K=h^{ij}k_{ij}=\frac{1}{2N}(h^{ij}h'_{ij}-2\partial_iN^{i}-h^{ij}N^m\partial_mh_{ij}).
\end{eqnarray}
Then the non-metricity scalar can be rewritten as
\begin{eqnarray}
	Q&=&-\frac{1}{4}Q_{\alpha\mu\nu}Q^{\alpha\mu\nu} +\frac{1}{2}Q_{\alpha\mu\nu}Q^{\mu\nu\alpha}+\frac{1}{4}Q_{\mu}Q^{\mu}-\frac{1}{2}Q_{\mu}\bar{Q}^{\mu}\nonumber\\
	&=& {}^{(3)}Q+K_{ij}K^{ij}-K^2-\frac{1}{N}K\partial_iN^{j}\nonumber\\
	&&+\frac{1}{N^2}\left[\partial_tN\left(\frac{\partial_iN^i}{N}\right)+\partial_tN^i\left(\frac{1}{2}h^{mn}\partial_ih_{mn}-\frac{\partial_iN}{N}\right)\right.\nonumber \\ 
	&&\left.\qquad+Nh^{ij}h^{mn}(\partial_jh_{mn}-\partial_mh_{nj})\partial_iN-\frac{1}{N}N^i\partial_iN\partial_mN^m+\partial_iN^j\left(\frac{1}{N}N^i\partial_jN-\partial_jN^i-\frac{1}{2}h^{mn}N^i\partial_jh_{mn}\right)\right].\nonumber\\
\end{eqnarray}
One can find that there are terms of time derivative of $N$ and $N_i$ in the form of non-metricity scalar, which leads to diffeomorphism symmetry breaking in $f(Q)$ theory.
For boundary term
\begin{eqnarray}
	B=\mathring{\nabla}_\mu(\bar{Q}^\mu-Q^{\mu})=\partial_\mu (\bar{Q}^\mu-Q^{\mu})+ \mathring{\Gamma}^{\sigma}{}_{\sigma\mu}(\bar{Q}^\mu-Q^{\mu}),
\end{eqnarray}
we can get
\begin{eqnarray}
	(Q-\bar{Q})^0&=&\frac{1}{N^2}(-h^{ij}\partial_th_{ij}+h^{ij}N^{m}\partial_mh_{ij}+\partial_iN^i)=-\frac{2K}{N}-\frac{\partial_iN^i}{N^2}\\
	(Q-\bar{Q})^i&=&\frac{N^i}{N}(2K+\frac{\partial_mN^m}{N})-\frac{1}{N^2}N^m\partial_mN^i+\frac{2}{N}h^{ij}\partial_jN+\frac{1}{N^2}\partial_tN^i+h^{ij}h^{mn}(\partial_jh_{mn}-\partial_nh_{mj}),
\end{eqnarray}
and
\begin{eqnarray}
	\mathring{\Gamma}^{\sigma}{}_{\sigma\mu}=\frac{1}{2}h^{ij}\partial_\sigma h_{ij}+\frac{\partial_\sigma N}{N}.
\end{eqnarray}
One can find that the boundary term contains second derivative of $h_{ij}$ and $N^i$.

\bibliographystyle{utphys}
\providecommand{\href}[2]{#2}\begingroup\raggedright\endgroup

\end{document}